\documentclass[preprint,showpacs,preprintnumbers,nofootinbib,amsmath,amssymb]{revtex4}
\usepackage{graphicx}
\usepackage{dcolumn}
\usepackage{bm}
\begin{document}
\title{On a test of the modified BCS theory performance\\ in the picket 
fence model}
  \author{Nguyen Dinh Dang$^{1, 2}$}
 \email{dang@riken.jp} 
\affiliation{1) Heavy-Ion Nuclear Physics Laboratory, RIKEN Nishina Center
for Accelerator-Based Science, 
2-1 Hirosawa, Wako City, 351-0198 Saitama, Japan\\
2) Institute for Nuclear Science and Technique, Hanoi, Vietnam}
\date{\today}
\begin{abstract}
The errors in the arguments, numerical results, and conclusions 
in the paper ``Test of a modified BCS theory performance in the picket 
fence model''~\cite{PV} by V.Yu. Ponomarev and A.I. Vdovin are pointed out.
Its repetitions of already published material are also discussed.
\end{abstract}

\pacs{21.60.-n, 21.10.Pc, 24.60.-k, 24.10.Pa}
\keywords{Suggested keywords}
\maketitle

\section{Introduction}
The modified BCS (MBCS) theory and its extension, the modified
Hartree-Fock-Bogoliubov (MHFB) theory, have been proposed and developed 
in a series of papers \cite{MBCS1,MBCS2,MHFB,MBCS3,MBCS4} between 2001
and 2007. The aim of these
theories is to restore the unitarity relation $R^{2} = R$ for the generalized
single-particle density matrix $R$, which is violated within the
conventional BCS and HFB theories at finite temperature $T$. 
This restoration leads to an application of the secondary Bogoliubov
transformation from the conventional quasiparticle operators to the
modified quasiparticle ones. As a result the MBCS equation is obtained.
The main merit of the MBCS theory is that it is a fully microscopic
theory, which shows that the phase transition from the superfluid phase to the normal
one (the SN phase transition) is smoothed out in finite systems. 
The MBCS pairing gap does not collapse at the critical temperature
$T_{c}$ of the SN phase transition, where the conventional BCS gap
vanishes. It also
points out, for the first time, that the origin of the smoothing out of the SN-phase
transition is the quasiparticle-number fluctuations (QNF), which are
ignored in the conventional BCS and HFB theories.
The MBCS theory has been applied with success in several open-shell realistic
nuclei~\cite{MBCS1,MBCS2,MHFB} 
and tested by its authors in an exactly solvable pairing model with $\Omega$
doubly folded levels~\cite{MBCS3,comment}, 
which is often referred to
as the Richardson model, ladder model, or picket-fence model (PFM).
Being an approximation that deals with the fluctuations due to the 
system finiteness, the MBCS theory is sensitive to the size of the system
under consideration. Within the half-filled PFM, e.g., i.e. when 
$N=\Omega$, where $N$ is the number of particles, the MBCS gap
decreases smoothly with increasing the temperature $T$ up to a certain
temperature $T_{M}$ ($> T_{c}$), where a discontinuity occurs (See
Refs. \cite{MBCS3,MBCS4}
and references therein). For a single-particle spectrum with the level
distance $\epsilon$ equal to 1 MeV, it has been found that 
his maximal temperature $T_{M}$ 
decreases almost linearly with decreasing $T$
from a value as high as $T_{M}\simeq$ 24 MeV 
for $N=\Omega=$ 100 to a value as low as $T_{M}\simeq$ 0.7 MeV 
for $N=\Omega=$ 6~\cite{MBCS3}. In Ref. \cite{comment} it has been
pointed out, for the
first time, that the
asymmetry of the QNF with respect to the Fermi energy is the reason that causes
the decrease of $T_{M}$ with $N$. In the same Ref.
\cite{comment}, it has also been found that
it is sufficient to add just one more level, i.e. $\Omega = N+1$, to
restore the symmetry of the QNF up to much higher $T_{M}$. This 
allows to extend the temperature region of applicability of 
the MBCS theory to a temperature as high as $T\leq$ 5 - 6 MeV for all
values of the particle number $N$ (for $\epsilon=$ 1 MeV and
the pairing interaction $G=$ 0.4 MeV). The MBCS theory has been the subject of exchanging 
criticisms between us~\cite{comment,MBCS3,MBCS4} and the authors of Refs. \cite{PV1,PV2}.
In a recent paper~\cite{PV}, the same authors 
reiterated their criticism by repeating the results of the 
same test within the PFM. 
Regrettably, beside largely duplicating the
contents, which have already been published in 
Refs. \cite{MBCS2,MHFB,MBCS3,comment,MBCS4,PV1,PV2}, the authors of Ref. \cite{PV} 
added nothing new but further serious mistakes and false conclusions. 
The aim of the present article is to 
point out the errors as well as
the repetitions in Ref. \cite{PV}, which is referred to hereafter as
the test \cite{PV}.
\section{Incorrect statements and erroneous results}
\subsection{Chemical potential}
\begin{figure}
    \includegraphics[width=12cm]{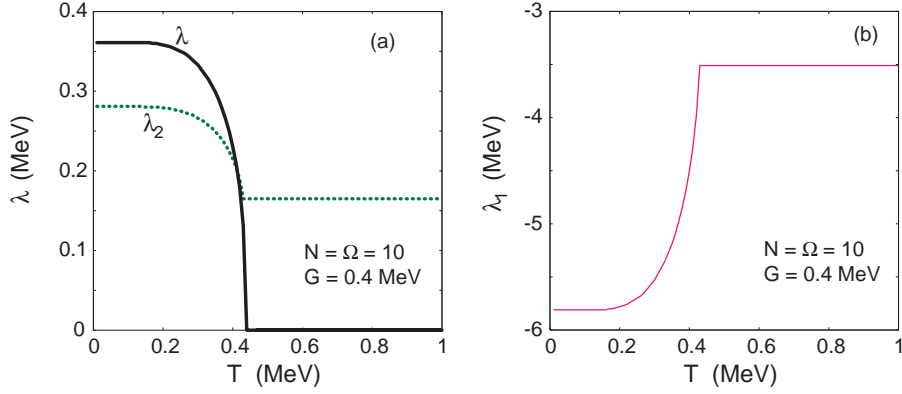} 
 \caption{(Color online) Chemical potentials $\lambda$, $\lambda_{1}$ 
 and $\lambda_{2}$ as functions of $T$ within the Lipkin-Nogami method for the
 PFM with $N=\Omega=$ 10 and $G=$ 0.4.\label{lambda}}
    \end{figure}

The authors of the test \cite{PV}
claim that the chemical potential of the PFM
should not change with temperature $T$, remaining always in the middle of the single-particle
spectrum taken to be zero in this case. 
This statement is true only within the BCS theory for the half-filled case (i.e.
$N=\Omega$) if the
self-energy correction term $\sim Gv_{j}^{2}$ is neglected in the
single-particle energies. Within the BCS theory at $T\neq$ 0,
the chemical potential $\lambda$ is determined as a Lagrangian
multiplier so that the
average particle number within 
the grand canonical ensemble (GCE) has the correct value. This is because the BCS theory violates
the particle number, causing quantal fluctuations of particle number, 
and ignores the effects due to the QNF at 
finite temperature as well. In any other theory, including the MBCS
theory, where these effects and/or correlations beyond the mean field 
are partially or fully taken into account, 
the chemical potential $\lambda$ becomes a
function of $T$. An example is the well-known Lipkin-Nogami (LN) method~\cite{LN}, which is an
approximate particle-number projection to partially eliminate the
particle-number fluctuations in the BCS theory. The LN method yields
$\lambda = \lambda_{1} + 2\lambda_{2}(N+1)$. These chemical potentials are 
shown in Fig. \ref{lambda} as functions of $T$ 
for the PFM with $N=\Omega=$ 10 and $G=$ 0.4 MeV. So long
as the pairing gap $\Delta$ exists, all of them change with $T$, 
and none of them remains at zero.

    \begin{figure}
        \includegraphics[width=14cm]{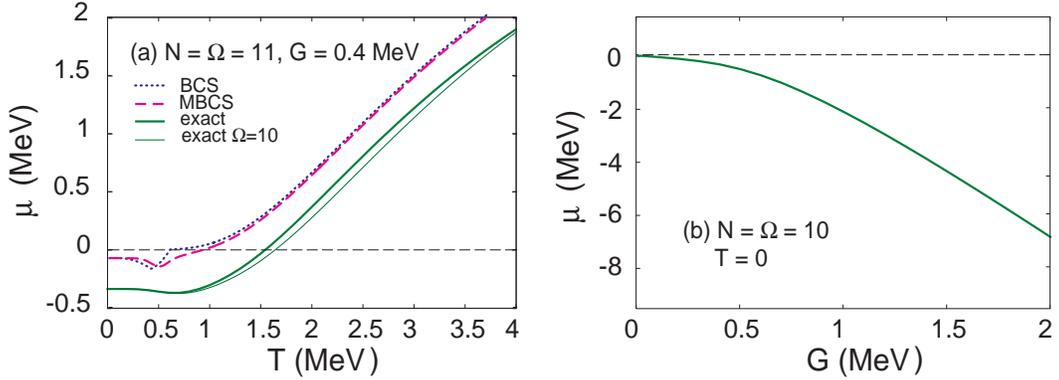} 
     \caption{(Color online) Chemical potentials $\mu$
     as a functions of $T$ at $G=$ 
     0.4 MeV (a), and a function of $G$ at $T=$ 0 (b) for the PFM with
     $\Omega = N+1=$ 11 (a), and $\Omega=N=$ 10 (b). The exact CE results
     are shown as the solid lines. In (a) the exact CE
     result for $\Omega=$ 10 is also shown as the thin solid line. The
     BCS and MBCS results are shown in (a) as the dotted and dashed lines,
     respectively.
     \label{muCE}}
        \end{figure}
The exact definition of the chemical potential $\mu$ is the difference
between the ground-state (internal) 
energies of the systems with $N\pm 2$ particles~\cite{Ring}, namely
\begin{equation}
\mu_{+} = \frac{1}{2}[{\cal E}(N+2) -{\cal E}(N)]~,\hspace{5mm} 
\mu_{-} = \frac{1}{2}[{\cal E}(N) -{\cal E}(N-2)]~,
\label{mu+-}
\end{equation}
\begin{equation}
\mu =\frac{1}{2}(\mu_{+}+\mu_{-}) = \frac{1}{4}[{\cal E}(N+2) -{\cal
E}(N-2)]~,
\label{mu}
\end{equation}
where ${\cal E}(N)$ denotes the ground-state energy of the
$N$-particle system in the zero-temperature case, or its internal
(total) energy 
at finite temperature $T$. In the latter case, the thermal average
within the GCE is carried out for a system which exchanges its energy and particle
number with a heat bath. If the particle number of the system 
is fixed, the canonical ensemble (CE) should be used for thermal averaging.
The PFM is a system with a fixed number of particles, where no particle
number fluctuations are possible. Therefore the meaningful 
exact results should be those obtained within the CE by using the
partition function constructed from the
exact eigenvalues of the pairing Hamiltonian [See Eqs.
(\ref{EGCE&ECE}) for
the expressions of the 
total energies within the GCE and CE]. 
As shown by the solid lines in Fig. \ref{muCE}, 
the chemical potential $\mu$ strongly depends on temperature $T$ and 
the interaction parameter $G$.
From Fig. \ref{muCE} (a) one can also see that the performance of the MBCS theory (the dashed line)
is quite reasonable.

    \begin{figure}
        \includegraphics[width=14cm]{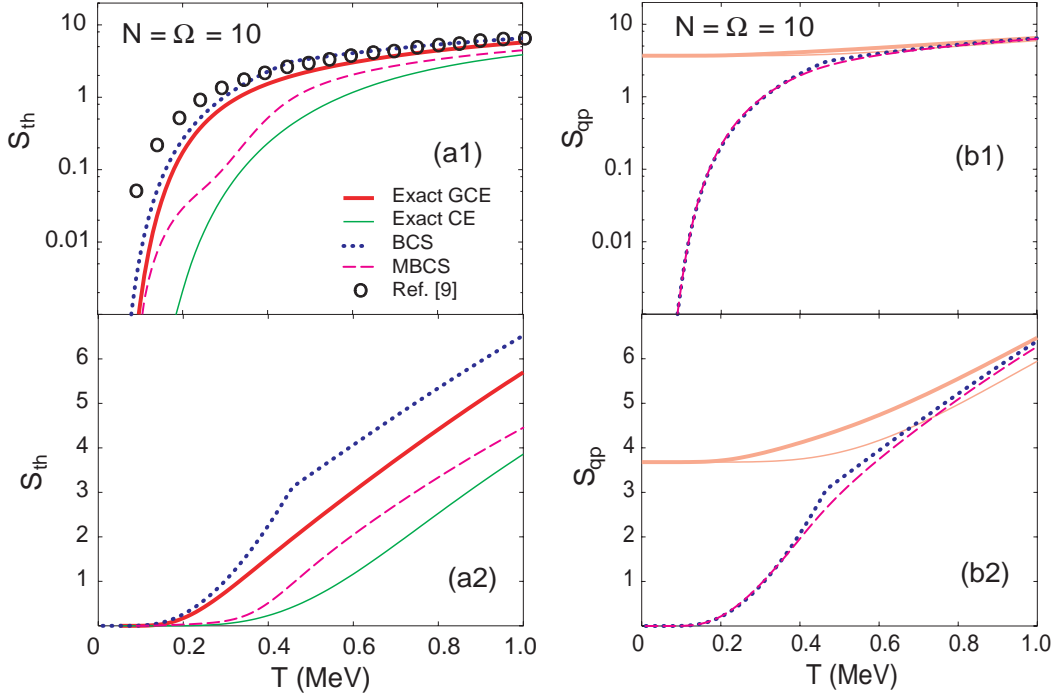} 
     \caption{(Color online) Entropies as functions of $T$ for
     $N=\Omega=$ 10 at $G=$ 0.4 MeV. The thermodynamic entropies $S_{th}$ are shown
     in panels (a), while panels (b) display the quasiparticle
     entropies $S_{qp}$.
     In the upper panels, (a1) and (b1), these quantities are shown in the
     logarithmic scale, whereas the linear scale is used in the lower
     panels, (a2) and (b2). The notations for the lines are shown in
     (a1), including the ``exact results'' of the test \cite{PV} (circles).
     In (b1) and (b2) the exact GCE and CE single-particle entropies
     are also shown as faint thick and thin solid lines, respectively.
     \label{S}}
        \end{figure}
\subsection{Entropies}
        Regarding the entropies, the results that the authors claim as the
exact thermodynamic entropy, denoted as circles in Fig. 7 (a) of Ref.
\cite{PV}, are incorrect. The same circles are also shown in Fig. \ref{S} 
(a1). The correct exact GCE and CE results for this quantity 
are shown as thick and thin solid lines, respectively,
in Fig. \ref{S} (a1) (in the logarithmic scale) and Fig. \ref{S} (a2) (in the
linear scale). They are obtained 
from the same Eq. (17) of the test \cite{PV} with the total energies
calculated within the GCE and CE by using the textbook
definitions, namely 
\begin{equation}
    \langle{\cal E}\rangle_{\rm GCE}=\frac{1}{{\cal Z}(\beta,\lambda)}\sum_{s,n}{\cal 
    E}_{s}^{(n)}e^{\beta\lambda n}p_{s}(\beta,n)~,\hspace{5mm}  
    \langle{\cal E}\rangle_{\rm CE}=\frac{1}{Z(\beta,N)}\sum_{s}{\cal 
        E}_{s}^{(N)}p_{s}(\beta,N)~,
        \label{EGCE&ECE}
        \end{equation}
        with the grand partition function ${\cal
        Z}(\beta,\lambda)$,
        and partition function $Z(\beta,n)$ defined as
        \begin{equation}
	   {\cal Z}(\beta,\lambda)=\sum_{n}e^{\beta\lambda n}Z(\beta, n)~,\hspace{2mm} 
	   {Z}(\beta, n)=\sum_{s}p_{s}(\beta,n)~,
	      \hspace{5mm} p_{s}(\beta,n)=d_{s}^{(n)}e^{-\beta{\cal
	      E}_{s}^{(n)}}~.
	   \label{Z}
	   \end{equation}
	   In Eq. (\ref{Z}), ${\cal E}_{s}^{(n)}$ are the eigenvalues 
	   of the pairing Hamiltonian for the system with $n$
	   particles, whereas $d_{s}^{(n)}$ are their degeneracies. 
	   The GCE sum in Eq. (\ref{EGCE&ECE}) is carried
	   out over all $n=$ 1, $\ldots$, $\Omega$-1 with blocking properly
	   taken into account for odd $n$.
As compared to these exact GCE values [thick solid line in Fig. \ref{S}
(a1)], the results of the test \cite{PV} (circles) at $T\leq$ 
0.1 MeV are wrong by more than one order of
magnitude. Meanwhile, the thermodynamic entropy obtained within the MBCS
theory (the dashed lines) are found sandwiched between the exact GCE and CE
results. The quasiparticle entropy obtained within the MBCS theory
agrees quite well with the BCS one, except for the region around the
critical temperature $T_{c}$, where the BCS gap collapses. At high $T$ 
they both converge to the exact GCE result for the single-particle
entropy as expected [See faint thick solid lines Figs. \ref{S}
(b1) and \ref{S} (b2)]. The exact GCE and CE single-particle entropies
are obtained by using the occupation
numbers $f_{j}$ on the $j$th single-particle orbital within the GCE
and CE, respectively. The latter are the
ensemble averages of the exact state-dependent occupation numbers
$f_{j}^{(s)}$, namely
\begin{equation}
    f_{j}^{(\rm
    GCE)}=\frac{1}{{\cal Z}(\beta,\lambda)}\sum_{s,n}f_{j}^{(s,n)}
    z_{s,n}(\beta,\lambda)~, \hspace{5mm} 
    f_{j}^{(\rm
       CE)}=\frac{1}{{Z}(\beta, N)}\sum_{s}f_{j}^{(s,N)}
       p_{s}(\beta, N)~,
       \label{fj}
       \end{equation}
       with
       \begin{equation}
    f_{j}^{(s)}=\frac{\sum_{k}N_{j}^{(k)}(C_{k}^{(s)})^{2}}
    {\sum_{k}(C_{k}^{(s)})^{2}}=\sum_{k}N_{j}^{(k)}(C_{k}^{(s)})^{2}~,
    \hspace{2mm} {\rm with}\hspace{2mm} \sum_{k}(C_{k}^{(s)})^{2}=1~,
\label{fjs}
\end{equation}
where $(C_{k}^{(s)})^{2}$ determine the weights of the eigenvector
components, and $N_{j}^{(k)}$ are the partial occupation numbers
weighted over the basis states $|k\rangle$.
\section{Repetitions of published material}
Sections 2 and 3 of the test \cite{PV} repeat 
the arguments of Ref. \cite{PV2}, where
the same authors replied to the our comments in Ref. \cite{comment}. 
This is clearly seen, e.g., in Fig. 1 of the test \cite{PV}, 
which is obtained by putting all the lines from the upper panels (for the pairing gaps) 
of Fig. 1 in Ref. \cite{PV2} into one panel. 
The only difference is that Fig. 1 of Ref. \cite{PV2} is for N = 10, while 
the same figure in the test \cite{PV} is presented for N = 14. 
This modification obviously offers no new physics as compared to Fig. 1 in
the previously published Ref. \cite{PV2}.

Section 3 of the test \cite{PV} also repeats the discussion of Refs.
\cite{MHFB,MBCS3,MBCS4,comment,PV1,PV2}. As we have pointed out in Refs.
\cite{MHFB,MBCS3,MBCS4,comment}, the validity of
the MBCS depends on how the QNF is taken into account. If the
single-particle spectrum is too small, the QNF can be large even at the
top and bottom of the spectrum, and its profile becomes strongly
asymmetric with respect to the middle of the spectrum already a low $T$.
This leads to some artifacts as negative values of the pairing gap, or 
a discontinuity in its temperature dependence. 
Therefore, a criterion is introduced to include the QNF symmetrically
from both sides of the Fermi energy in such a way that the negative and positive wings discussed in Fig. 2 (b) of Ref.
\cite{MHFB} nearly cancel each other. For the PFM under consideration, it turns out that, for
small $N$, e.g. $N\leq$ 14 with the level distance equal to 1 MeV and G= 0.4 MeV, it is sufficient to
enlarge the space by just one level $\Omega = N+1$ to achieve a symmetric profile of the QNF up to a
rather high temperature, while adding or subtracting more levels again decreases the value of the
limiting temperature, at which the QNF profile becomes asymmetric. For realistic heavy
nuclei, such as $^{120}$Sn e.g., this criterion is often unnecessary so long as 
the whole single-particle space is included
(Fig. 1 - f of Ref. \cite{MHFB}). In this respect, Figs. 2 - 4 of the 
test \cite{PV} and the related discussions therein are just a disguised tautology
since they demonstrate nothing but the asymmetry of the QNF, which has already
been shown and discussed previously in Fig. 2 of Ref.~\cite{MBCS3}, and Fig. 1 of 
Ref.~\cite{MBCS4}.

In Fig. 2 (b) of our paper \cite{MHFB}, the partial components of the thermal gap 
are shown to clarify the temperature 
dependence of the total gap, which is a physical quantity.
It seems that the authors of the test \cite{PV} misunderstood this idea
when they borrowed our results to divide the MBCS thermal gap into
the hole and particle gaps [See Eqs. (15) and (16) of the test \cite{PV}].
They summed up the values of $\delta\Delta_{j}$ discussed in 
our paper \cite{MHFB} over the levels below the
Fermi energy separately from those above it, and considered these two sums as 
the hole and particle
gaps, respectively. Obviously, such separation
of the thermal gap into the hole and particle gaps as two physical quantities
is misleading. The MBCS thermal gap is level-independent, and 
induced by all the QNF, which should be included by summation
over all quasiparticle orbitals. Figures 5 (c) and 5 (d) of the test \cite{PV} 
repeat the upper panel of Fig. 3 in
Ref. \cite{PV1}, which has previously been published by the same
authors under almost the same title. Two additional panels (a) and (b)
bear no additional physics information. These issues have been refuted
by us in Ref. \cite{comment}. 

Last but not least, that the MBCS
theory does not cure the particle-number fluctuation (PNF) 
has already been discussed long ago in Ref. \cite{MBCS2}, where the
PNF was calculated for several neutron-rich Ni isotopes~\cite{MBCS2}. 
This is the reason why several methods of particle-number
projection were applied within the MBCS
theory in Ref. \cite{MBCS4}, whose calculations were carried out not only for
the PFM but also for the realistic nucleus $^{120}$Sn. 
Therefore, Fig. 6 and the related
discussions in the test \cite{PV} are redundant because they just 
reiterate a well-known fact by using an oversimplified toy model with
just two levels.
\section{Introducing misleading quantities and making wrong
conclusions}
A part from the errors and repetitions mentioned above,  
the authors of the test \cite{PV} also introduced misleading discussions by
comparing the MBCS predictions of quasiparticle energies with the
``exact" ones (See Fig. 3 of the test \cite{PV}), whose definition remains
completely obscure. A quasiparticle state is an approximation, 
which arises from the canonical Bogoliubov transformation, and is related
to the mean-field concept.
So long there is no exact mean field, there cannot be exact quasiparticle 
energies.
The eigenvalues of the PFM, which are obtained by diagonalizing the
pairing Hamiltonian or solving the Richardson's equations, are not quasiparticle
energies. In order to have a better agreement with the
exact eigenvalues one needs to go beyond the conventional BCS theory
to take into account the correlations beyond the mean field. It is for
this purpose that the particle-particle
self-consistent RPA (SCRPA)~\cite{ppRPA1,ppRPA2} (for low value of the pairing
interaction parameter $G$) and quasiparticle RPA
(SCQRPA)~\cite{SCQRPA} (for an arbitrary  value of $G$) were developed.
But while the SCRPA and SCQRPA can describe
rather well the low excitation energies, they cannot describe the complicate
splitting of high-lying levels due to configuration mixing in the exact
solutions~\cite{Yuz}.

\begin{figure}
    \includegraphics[width=14cm]{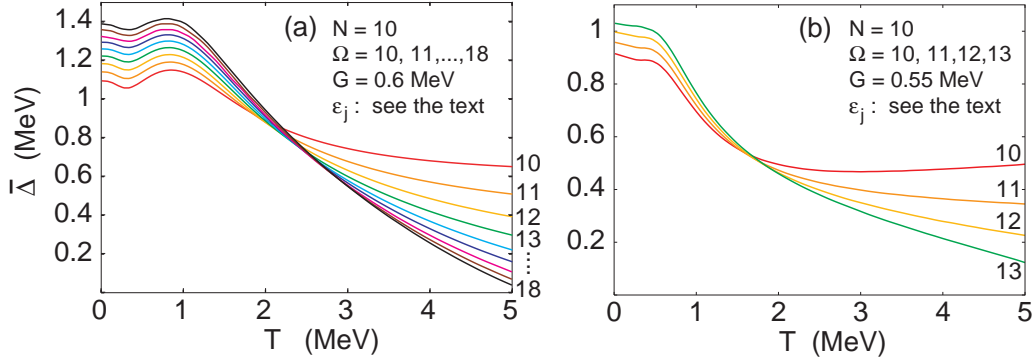} 
 \caption{(Color online) MBCS gaps $\bar{\Delta}$ as functions of temperature $T$
 obtained within the PFM, which has $N=$ 10 particles, and $\Omega=N+k$ levels. 
 (a) Results obtained with $G =$ 0.6 MeV, 
 $k=$ 0, 1,\ldots, 8;  (b)  
 Results obtained with $G =$ 0.55 MeV, 
 $k=$ 0, 1, 2, and 3. The numbers at the right margins of (a) and (b) 
 show the corresponding values of $\Omega$. The 
 single-particle energies for the sets (a) and (b) are 
 specified in the text.
 \label{contre}}
    \end{figure}
Apart from the conclusions of the test \cite{PV}, which 
obviously become invalid as a consequence of the above analysis, 
the authors of the test \cite{PV}
also claim: ``{\it We confirm that there exists a single example of the PFM 
(with the number of levels $\Omega$ equal
to the number of particles $N$ plus one) 
in which the MBCS produces the thermal behavior of
the pairing gap similar to the one of a macroscopic 
theory up to rather high temperatures. On
the other hand, we demonstrate that in all other examples of the PFM with 
$\Omega\neq N$ the theory
predicts phase transitions of unknown types at a much lower temperature.}'' 
Of course, this statement is
incorrect. It is valid only for 
a PFM with $N=$ 10, $G=$ 0.4 MeV with the level distance $\epsilon=$ 1 MeV.
It is not difficult to find counter examples, some of which are 
shown in Fig. \ref{contre}. It depicts the MBCS
pairing gaps $\bar{\Delta}$ obtained for the PFM with $N=$ 10, and
$\Omega=N+k$, which are grouped in two sets. The set in Fig. \ref{contre} (a)
consists of nine MBCS gaps, which are obtained with $k =$ 0, \ldots, 8 
by using $G=$ 0.6 MeV, and the 
following single-particle energies: $\epsilon_{j}=2.1\times j$ (MeV)
for $j\leq$ 3 and $j\geq$ 7, whereas $\epsilon_{j}=\epsilon_{3}+j-3$
(MeV) for 4$\leq j\leq$ 6.
The set in Fig. \ref{contre} (b) represents four MBCS gaps, which are
obtained with $k=$ 0, 1, 2, and 3 by using $G=$ 0.55 MeV, and 
the following single-particle energies: 
$\epsilon_{j}=2.1\times(j-5.5)$ (MeV) for $j\leq$ 3 and 7 $\leq j\leq$ 10, 
$\epsilon_{j}=j-5.5$ (MeV) for 4 $\leq j\leq$ 6, and 
$\epsilon_{j}=2.1\times(0.55\times\epsilon_{10}+j-10)$ (MeV) 
for 11 $\leq j\leq$ 13. All thirteen MBCS gaps in Fig.
\ref{contre} are smooth functions of $T$ at 0 $\leq
T\leq$ 5 MeV. This figure clearly shows how changing the size of the
system affects the tail of the gap at $T\geq$ 2 MeV in these cases.
These examples and the analysis in the present article 
are more than sufficient to rule
out all the conclusions of the test \cite{PV}.
\section{Conclusion}
In conclusion, the test \cite{PV} not only repeats already published
results, but also contains incorrect statements and detectable errors,
which make its conclusions invalid.
\acknowledgements
I thank N. Quang Hung for critically reading the manuscript and
assistance in numerical calculations, which 
were carried out by using the {\scriptsize FORTRAN} IMSL
Library by Visual Numerics on the RIKEN Super Combined Cluster
(RSCC) system. 

\end{document}